\documentclass{aastex}
\usepackage{spr-astr-addons}

\def\mbh{$M_{\rm BH}$\/}
\def\nh{$n_{\mathrm{H}}$\/}
\def\lledd{$L/L_{\rm Edd}$}

\def\rfe{$R_{\rm FeII}$}

\def\msol{M$_\odot$\/}

\def\rg{R$_{\rm g}$\/}
\def\ltsima{$\; \buildrel < \over \sim \;$}
\def\ltsim{\lower.5ex\hbox{\ltsima}}  
\def\simlt{\lower.5ex\hbox{\ltsima}}  
\def\gtsima{$\; \buildrel > \over \sim \;$}

\def\gtsim{\lower.5ex\hbox{\gtsima}} 
\def\simgt{\lower.5ex\hbox{\gtsima}}

\def\ha{{\sc H}$\alpha$}
\def\lya{{ Ly}$\alpha$}
\def\civ{{\sc{Civ}} $\lambda$1549\/}
\def\civnc{{\sc{Civ}} $\lambda$1549$_{\rm NC}$\/}

\def\cmq{cm$^{-2}$\/}
\def\cm3{cm$^{-3}$\/}
\def\hb{{\sc{H}}$\beta$\/}

\def\mgii{{Mg\sc{ii}} $\lambda$2800\/}

\def\oiiiopt{{\sc{[Oiii]}}\- $\lambda\lambda$\-4959,\-5007\/}
\def\o4363{{\sc{[Oiii]}} $\lambda$4363\/}
\def\caii{{Ca{\sc ii}}}

\def\heiiuv{He{\sc{ii}} $\lambda$1640}

\def\feiiopt{{Fe \sc{ii}}$_{\rm opt}$\/}
\def\feii{{Fe\sc{ii}}\/}

\def\nii{{[N\sc{ii}]} $\lambda\lambda$6548,6583\/}
\def\sii{{S\sc{ii}} $\lambda\lambda$6717,6730\/}
\def\oii{[{O\sc{ii}]} $\lambda$3727\/}

\def\fe{{\sc{Fe}}\/}

\def\fe76087{{\sc [Fe vii]}$\lambda$6087\/}
\def\oiii{{\sc [Oiii]}$\lambda$5007}

\def\kms{km~s$^{-1}$}

\def\ergss{ergs s$^{-1}$\/}

\def\hi{H{\sc i}\/}

\def\heii{{{\sc H}e{\sc ii}}$\lambda$4686\/}

\def\ovi{O{\sc vi } $\lambda$1035\/}

\def\oi{O{\sc i} $\lambda$1304\/}

\RequirePackage{color}

\begin{document}

\title{The most powerful quasar outflows as revealed by the \civ\ resonance line$^{1}$}

\shorttitle{Short article title}
\shortauthors{Autors et al.}

\author{P. Marziani}
\affil{INAF, Osservatorio Astronomico di Padova, Padova, Italia}
\and 
\author{M. A. Mart\'{\i}nez Carballo}
\affil{Instituto de Astrof{\'\i}sica de Andaluc{\'\i}a (CSIC),    Granada, Spain}
\and
\author{J. W. Sulentic}
\affil{Instituto de Astrof{\'\i}sica de Andaluc{\'\i}a (CSIC),    Granada, Spain}
\and
\author{A. Del Olmo}
\affil{Instituto de Astrof{\'\i}sica de Andaluc{\'\i}a (CSIC),    Granada, Spain}
\author{G. M. Stirpe}
\affil{INAF, Osservatorio Astronomico di Bologna, Bologna, Italia}
\and
\author{D. Dultzin}
\affil{Instituto de Astronom{\'\i}a, UNAM, 
Mexico, D.F.,   Mexico}
 

\altaffiltext{1}{Based in part on observations made with ESO Telescopes at the   Paranal Observatory under programme 082.B- 0572(A), and on observations made with the Italian Telescopio Nazionale Galileo (TNG). }

\defcitealias{zamanovetal02}{Z02}

\begin{abstract}
While quasar outflows may be quasi-\-ubi\-qui\-tous, there are significant
differences on a source-by-source basis. These differences can be
organized along the 4D Eigenvector 1 sequence: at least at low $z$, with only Population A sources radiating at relatively high Eddington ratio  and showing prominent high-velocity
outflows in \civ\  line profiles. We discuss in this paper VLT-FORS observations of \civ\ emission line profiles for a high-luminosity sample of Hamburg-ESO quasars and how they are affected by
outflow motion as a function of quasar luminosity. Our high-luminosity sample has the
notable advantage that the rest frame has been accurately determined from
previous VLT-ISAAC observations of H$\beta$\ in the J, H, and K bands. This makes measures of
inter-line velocity shifts  accurate and free of systemic biases. As the 
redshift increases  and the luminosity of the brightest quasars increases, powerful, high-velocity
outflows become more frequent.  We discuss the outflow contextualisation,
following the 4DE1 formalism, as a tool  for understanding the nature of  the so-called weak
lined quasars (WLQ) discovered  in recent years as a new, poorly
understood class of quasars.
We estimate the kinetic power associated with  \civ\ outflows and suggest
that the host galaxies in the most luminous sources likely experience
significant feedback.
\end{abstract}


\section{Introduction}
\label{intro}

The  optical and UV spectra of quasars show  broad and narrow lines
emitted by ionic species over a wide range of ionization potentials.
A quasar spectrum is usually easily recognizable with the same emission
lines almost always present  and usually strong ($W \sim $ 20 -- 200 \AA)
(with some exceptions   \S \ref{wlq})  in all type-1 AGN.  It is useful to
distinguish between high- and low-ionization lines (hereafter HILs and LILs, respectively): \civ, \heii, \heiiuv,
\ovi\ among the strongest HILs.  Balmer (\hb, \ha), optical \feii, \mgii
and the \caii\ IR triplet are frequently observed LILs. We can also
between high
and low ionization narrow lines: \oiiiopt, \heiiuv, \heii, [Ne{\sc iii}] 
(HILs) and  Balmer, \oi, \sii, \nii\ (LILs). Apart from these commonalities, it is important to remark that quasars show widely differing line profiles, intensity ratios, and ionization levels (e.g., \citealt{marzianietal15}). This is made evident by a comparison between the prototypical Narrow Line Seyfert 1 (NLsy1) I Zw 1,  and a broader line object like NGC 5548 \citep{sulenticetal00a}. In this context it is especially important to stress that  lines do not all show the same profiles. Internal emission line shifts involve both broad and narrow lines \citep{zamanovetal02,eracleoushalpern03,huetal08,hewettwild10}. Narrow LILs such as narrow \hb\ and \oii\ are the most suitable for estimating the quasar systemic redshift; however they are not easily observable at  $z>1$ since they are shifted into the IR.  Quasar systemic redshifts reported in literature are significantly more uncertain if $z>1$. Systemic biases in the SDSS spectra as large as 600 \kms\ have been identified \citep{hewettwild10}. They are of the same order of magnitude  of internal emission line shifts between HILs and LILs in quasars \citep{gaskell82,tytlerfan92,marzianietal96,corbin97}.

There is general consensus that  broadening of optical and UV emission lines is mostly associated with  the Doppler effect reflecting emitting gas motions relative to the observer  (scattering and gravitational redshift effects are not clearly assessed in quasar spectra e.g.,  \citealt{gaskellgoosmann13}).  HIL blueshifts provide evidence of outflow (if the receding side of the outflow is hidden from view, as in an optically thick accretion disk with associated wind) but require rest frame knowledge and proper contextualisation for a correct analysis. A first contextualisation (e.g. radio quiet vs. radio loud;
\citealt{sulenticetal95,corbin97}) showed intriguing differences
between RQ and RL quasars with large blueshifts observed only among RQ 
sources (\S \ref{4de1}). We discuss here our VLT-FORS \civ\ spectra of
high luminosity quasars  (their location in the absolute magnitude vs. redshift plane is shown in Fig. \ref{fig:sample})   where  a reliable rest frame could be estimated
from narrow LILs, typically the narrow component of \hb\ (\S \ref{obs}). 
We report on the
behaviour of the \civ\ emission line profile as a function of luminosity,
using the interpretative tools provided by the 4D eigenvector 1  approach (4DE1, \S
\ref{res}). We
derive inferences about the nature of weak-lined quasars (\S \ref{wlq})
and demonstrate that nuclear outflows traced by \civ\ are expected to have
a
strong feedback effect on the host galaxy, especially for the most
luminous Pop. A quasars (\S \ref{out}).


\section{(4D)Eigenvector 1 quasar contextualization at low-$z$}
\label{4de1}

The (4D) Eigenvector 1 quasar con\-text\-ualiz\-ation sch\-eme at low-$z$
offers a powerful tool for
interpretation of the \civ\ emission line profile.  Eigenvector 1 was
originally defined from a Principal Component Analysis
(PCA) of 87 PG quasars involving an anticorrelation between optical FeII
intensity, half-maximum profile width of \hb\  and peak intensity of \oiii\ \citep{borosongreen92}.  E1 expanded to 4DE1
with the addition of X-ray photon index and \civ\ profile
shift measures \citep{sulenticetal00a,sulenticetal08,sulenticetal11}. 
4DE1 allows one to define a quasar main sequence in 4DE1 space
\citep{sulenticetal00a,marzianietal01}. The 4DE1 approach allows the
definition of spectral types following source occupation in the optical
plane FWHM(H$\beta$) vs \rfe\ \citep{sulenticetal02} and can be extended
to include all four dimensions e.g. \citep{bachevetal04}.
Along the sequence, it is possible to identify two populations: Population
A with FWHM(H$\beta$)$\le 4000$ \kms, and a second Population B of sources with broader lines \citep[see ][for the rationale behind two distinct quasar populations]{sulenticetal07,sulenticetal11}. Here we mention that, at low-$z$\ ($\ltsim 1.0$),  Pop. A sources include narrow-line Seyfert 1s (NLSy1s) and strong \feiiopt\ emitters. Pop. B sources show weaker \feii\ emission and LIL emission broad profiles that are often redward asymmetric \citep{marzianietal03a}.  In low-$z$\ samples Pop. B sources are systematically  more massive (\mbh $\sim 10^{9}$ \msol) than Pop. A, and the wide majority of Fanaroff-Riley II powerful radio sources belongs to Pop. B \citep{zamfiretal08}. Pop. B sources could be therefore seen as  more ``evolved''  AGN than Pop. A sources.  Heuristic considerations and black hole and Eddington ratio estimates suggest that the principal  driver of the quasar main sequence is Eddington ratio
\citep[e.g.,][]{marzianietal01,marzianietal03b,kuraszkiewiczetal09}. 
Population A and B are distinct in terms of
Eddington ratio: the boundary value (for BH mass of 10$^8$ \msol) is
estimated to be \lledd $\approx 0.2 \pm 0.1$ \citep{marzianietal03b}.
It is interesting to note that this is the limit at which the transition
from a geometrically thin to a geometrically thick disk
is expected \cite[][and references therein]{abramowiczetal88,franketal02}.

\subsection{Interpretation of the \civ\ emission line profile along the 4DE1 sequence}

The strongest emission lines   can be empirically reproduced by 3 components with varying relative strength along the 4DE1 sequence \citep{marzianietal10}:  (a) a blueshifted component (BLUE), strong in \lya, \civ, \heiiuv;  (b) a ``Broad Component'' (BC), strong in all low ionization lines: \feiiopt, \mgii, \hi \ Balmer lines; (c) a ``Very Broad Component'' (VBC FWHM$\sim$10000 \kms; \citealt{petersonferland86,marzianisulentic93,corbin97,sulenticetal00b}), redshifted; strong in \lya, \civ, \heii, Balmer lines but  absent or at most weak in \feii.

Blueshifts have been associated with radial motion + obscuration since their discovery \citep{gaskell82}. The blueshifted component is meant to isolate, in a  heuristic way, the contribution of the radially-moving (most likely outflowing) gas. The approach is heuristic since, in a multicomponent decomposition, we are isolating radial velocity components that may have no obvious counterpart in physical space. At the same time,    inter-percentile velocity intensity ratios for the blue part of the profile   are very different from the ones estimated over the core or of lines showing an almost symmetric profile. Since the emission line profiles originate in a spatially unresolved region, to ignore emission line shifts and asymmetries, and take an intensity ratio over the full profile can easily bring to   misleading results.  We can use Mark 478 as a representative example of spectral type A2, the most populated A spectral type; the same consideration apply to BLUE of the other A spectral type \citep{marzianietal10}.  If the \civ/\hb\ ratio is computed over the full profile from the Mark 478 spectrum, it is $\approx 19$, for the unshifted BC component is $\approx$1.4.  Clear constraints on physical conditions emerge if the line components are considered individually.  It is possible to measure fairly accurately the following intensity ratios for BLUE: \lya/\hb\ ($\gtsim 30$), \civ/\lya\ ($\sim 0.5$), \heiiuv/\civ\ ($\gtsim 0.15$).  Results of Cloudy 08.00 \citep{ferlandetal98,ferlandetal13} simulations  as a function of density and ionization parameter $U $ \ indicate that  BLUE is consistent   with high ionization ($U \sim 10^{-1\pm0.5}$) and moderate density (\nh $\sim~10^{9.5\pm0.5}$ \cmq). The large \lya/\hb $\gtsim 30$\ is very different from the ones of the other components (\lya/\hb\ $\approx$ 5 -- 10, \citealt{netzeretal95}). This latter  value is  difficult to explain for photoionization models.  On the converse \lya/\hb\ $\gtsim 30$\ is expected under more standard physical conditions. 

The decomposition involving BLUE is   consistent with some models proposed in the past, such as those by \citet{collinsouffrinetal86} and \citet{elvis00}. In the framework of these models, the blueshifted component is associated with radially outflowing clouds surrounding an accretion disk, or with a (non-rotating) wind driven by radiation pressure. There is fairly convincing evidence that strongest outflows are associated with the highest Eddington ratio values (\lledd; e.g., \citealt{boroson02,marzianietal03b,baskinlaor05b,richards12,marzianietal13}) where we indeed expect the strongest effect of radiation pressure. 

Further  interpretation of the heuristic decomposition of the broad profiles involves ``stratification'' of the emitting region: the broad component is associated with a lower ionization Broad Line Region (BLR), where line broadening is predominantly virial,  and where \feii, \caii\ are emitted \citep[][and references therein]{martinez-aldamaetal15}.  The very broad component is associated with a high-ionization inner region, the Very Broad Line Region (VBLR) that emits no \feii\ and shows lower continuum responsivity \citep{sneddengaskell07,goadkorista14}. The stratification is consistent with  virial motions dominating the core of the Pop. B LILs. The LIL profiles often show asymmetries but the amplitude of the shifts is usually much less than the FWHM \citep{sulentic89,marzianietal03a,zamfiretal10}. For reviews on the relation between emission line properties and  accretion disk see \citet{gaskell08} and \citet{gaskell09b}.

\subsection{Interpretation of the line profile: a narrow component for \civ?}

The long-slit spectra are far from resolving the broad line emitting region (size $\sim 1 - 10$ pc). The  half-slit width of 0.3 arcsec covers $\approx \pm$ 2.5\ kpc  from the nucleus. In addition to the BLR, a part of the (usually resolved) narrow line region (NLR) is included in the slit. The profiles of \civ\ and H$\beta$\ are therefore the summation of contributions which may be separated (or overlapping) in radial velocity, but not necessarily spatially disjoint (or coincident). The interpretation of the line profile will follow a series of heuristic considerations which come from past analysis of large samples of high S/N spectra.  As mentioned, we identified 4 main components: 1) the narrow component; 2) the broad component; 3) the very broad component; 4) a blueshifted component \citep{marzianietal03b,marzianietal10}. The NC is partially resolved in nearby AGN. It is easily appreciable in the H$\beta$\ profile of Pop. B sources. Its presence has been debated for \civ, several convincing arguments notwithstanding \citep{sulenticmarziani99}. \citet{sulenticmarziani99} ascribed the \civ\ NC to the innermost, densest part of the NLR.  Empirical isolation of the \civnc\ is often ambiguous, as it merges smoothly with the underlying BC. While its contribution to the total \civ\ flux is small, it can significantly affect line width measures. However, the very high luminosity of the HE sample may imply that the NC is contributing little to both BC flux and width \citep{netzeretal04}. 

\subsection{Modelling and measuring the broad \civ\ profile}

\begin{figure*}[htp]
\includegraphics[scale=0.475]{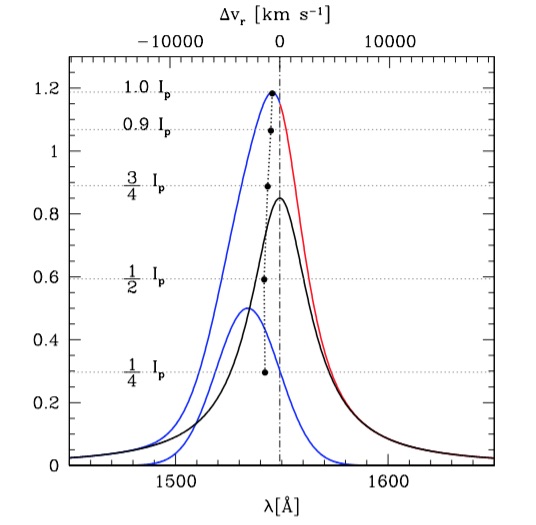}
\includegraphics[scale=0.475]{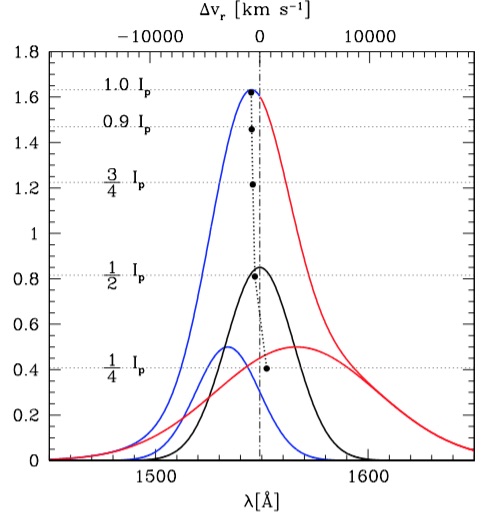}
\caption{Mock \civ\ profile illustrating the line decomposition into a BC and a blueshifted  component for Pop. A (left), and into a BC, VBC and blue shifted component for Pop. B (right). Radial velocities are measured for the peak intensity $I_\mathrm{p}$,  and the centroids (black spots) at different fractional intensities on the {\em full} profile, to obtain a quantitative parameterisation that is independent of the decomposition of the profile.   }
\label{fig:mock}        
\end{figure*}

Fig. \ref{fig:mock} illustrates   the \civ\ line decomposition into a BC and a blueshifted component, for Pop. A, and into BC, VBC and blueshifted component for Pop. B  (right panel). Radial velocities are measured for the peak intensity $I_\mathrm{p}$,  and for the centroids (black spots) at different fractional intensities. The same decomposition can be applied also to \hb, although the blue shifted component is usually close to or below the noise level. Line centroids at fractional intensity are independent of multi-component decomposition, and will be considered for a quantitative statistical analysis in the following discussion.

\begin{figure}[htp]
\includegraphics[width=0.465\textwidth]{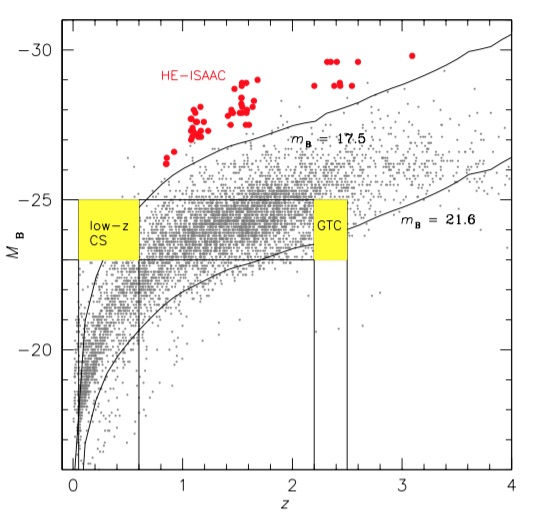}
\caption{Absolute B magnitude vs $z$ for the HE sample with ISAAC observations (red dots). A random subsample of the \citet{veroncettyveron10} catalog is also plotted (grey dots). The absolute magnitude associated with two flux limits (17.5, apropriate for the HE survey) and (21.6, appropriate for the SDSS) are shown as function of redshift. Regions representing a low-$z$ control sample and a GTC sample of faint quasars at intermediate $z$\  (as described in \citealt{sulenticetal14}) are shown shaded in yellow.}
\label{fig:sample}       
\end{figure}

\section{Sample selection, observations and data analysis}
\label{obs}

Our sample was drawn from 52  quasars from the Hamburg-ESO survey  \citep{wisotzkietal00} observed with ISAAC in the \hb\ spectral range to  obtain reliable rest frames,  in the redshift range $0.9 \lesssim z \lesssim 3$ (most with $ z \lesssim 3$).   The \civ\ line was observable from ground ( $z \gtsim 1.4$) for 32 sources, and \civ\ was actually observed  at TNG (equipped with DOLORES) and VLT (equipped with FORS2) for  28 quasars (hereafter the HE sample). The quasar HE  sample includes 15 Pop. A  and 17 Pop.  B sources.  A control sample (CS) for which \hb\ and \civ\ observations are available at low $z$ and low luminosity was extracted from 130 HST/FOS \civ\ observations \citep{sulenticetal07}. Fig. \ref{fig:sample} illustrates the placement of the CS and of the HE sample in the absolute magnitude vs. redshift plane. In this paper we will use all of the 130 FOS observations, which means the addition of a minority of sources more luminous than $M_\mathrm{B} \approx -25$. An additional sample of GTC observations of faint quasars at high-$z$\ has been also used, to help distinguish redshift from $L$\ effects.  

Previous observations of H$\beta$\ for the HE sample were discussed in a series of papers  \    \citep{sulenticetal04,sulenticetal06,marzianietal09}. Data analysis   has been carried out with multicomponent maximum-likelihood  fits in both the UV and optical rest-frame range. The IRAF task SPECFIT \citep{kriss94} allows for the inclusion of most-frequently observed components in the spectra of extragalactic sources. Specific to the analysis of \civ\ and H$\beta$\ in quasars, we considered a local power-law, \feii\ emission represented by a scaled and broadened template. Narrow lines were fit with Gaussians, while broad \civ\ and H$\beta$\ were fit using Lorentzian and/or Gaussians, depending on Population as described below. 

\begin{figure*}[htp!]
\includegraphics[width=0.475\textwidth]{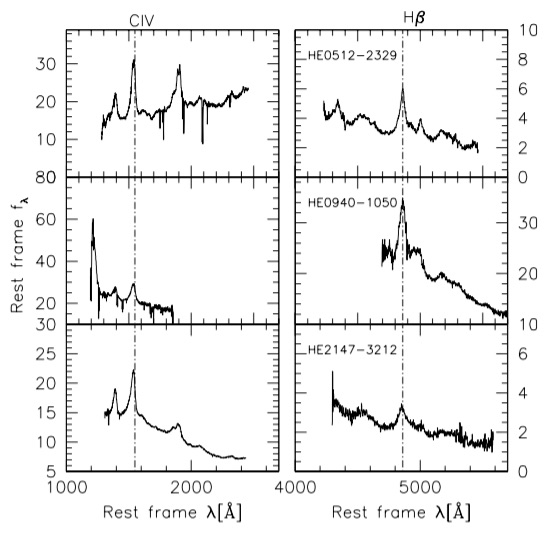}
\includegraphics[width=0.475\textwidth]{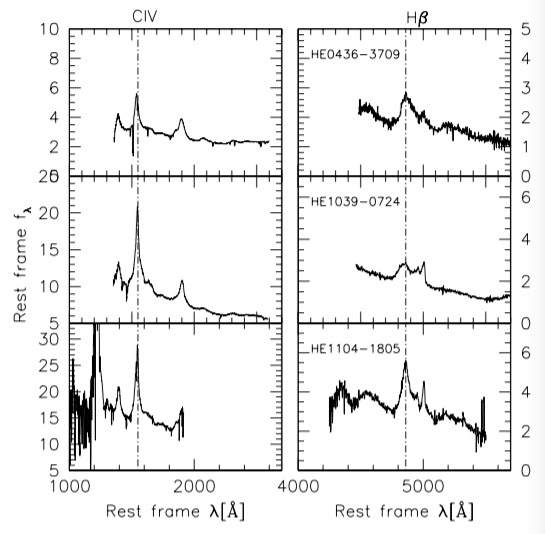}
\caption{Spectra of  sources in the HE sample, in the \civ\ and \hb\ spectral range, for Pop. A (left) and Pop. B (right). Absciss\ae\ are rest frame wavelength in \AA, ordinates are specific flux in units of  $10^{-15}$ \ergss\ \cmq\ \AA$^{-1}$. }
\label{fig:spectra}       
\end{figure*}

\section{Results}
\label{res}

\begin{figure*}[htp!]
\includegraphics[width=0.455\textwidth]{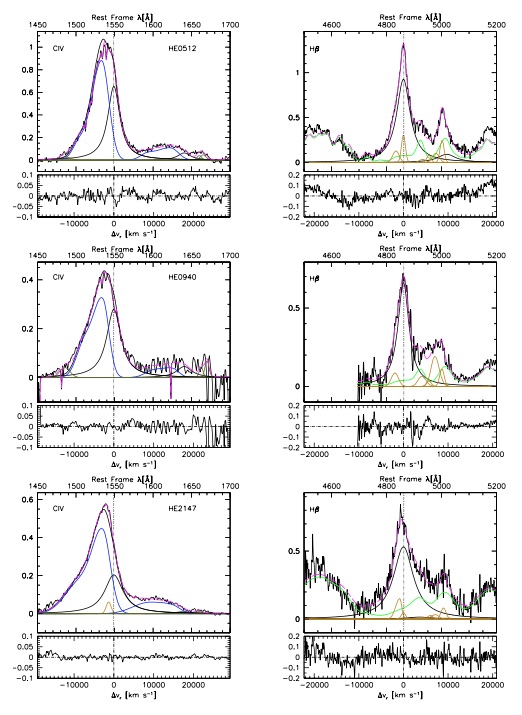}
\includegraphics[width=0.467\textwidth]{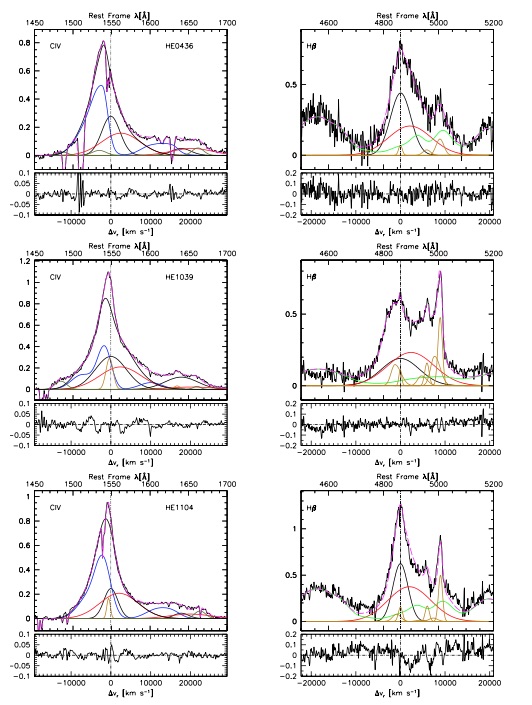}
\caption{Spectra of  sources in the HE sample, in the \civ\ and \hb\ spectral range, for Pop. A (left) and Pop. B (right), after continuum subtraction.   Horizontal scale is rest frame wavelength in \AA\  or radial velocity shift from \civ\ rest wavelength (left) or  \hb\ (right) marked by dot dashed lines; vertical scale is  intensity normalized to specific flux at continuum adjacent the two emission lines.  The panels show the emission line components used in the fit:  \feii\ emission (green),    broad and very broad component (red, for Pop. B only), and BLUE (blue).   Lower panels show residuals observed - SPECFIT model.  }
\label{fig:profiles}       
\end{figure*}

Fig. \ref{fig:spectra} provides   examples of Pop. A  and Pop. B sources in the HE sample, with the \hb\ and \civ\ spectral ranges shown side-by-side. The result of the SPECFIT analysis carried out for these objects is shown in Fig. \ref{fig:profiles}. A most remarkable finding is that the \civ\ profiles show significant blueshifts with respect to the rest frame (dot-dashed line) for both Pop. A and B. The amplitude of the blueshifts are however larger for Pop. A than for Pop. B. A detailed analysis of the full dataset will be presented elsewere. Here we focus on results concerning the relation between blueshift and quasar luminosity.

\subsection{Luminosity and radio-loudness effects on quasars blueshifts}
\label{lumrl}

\begin{figure*}[htp!]
\includegraphics[scale=0.21]{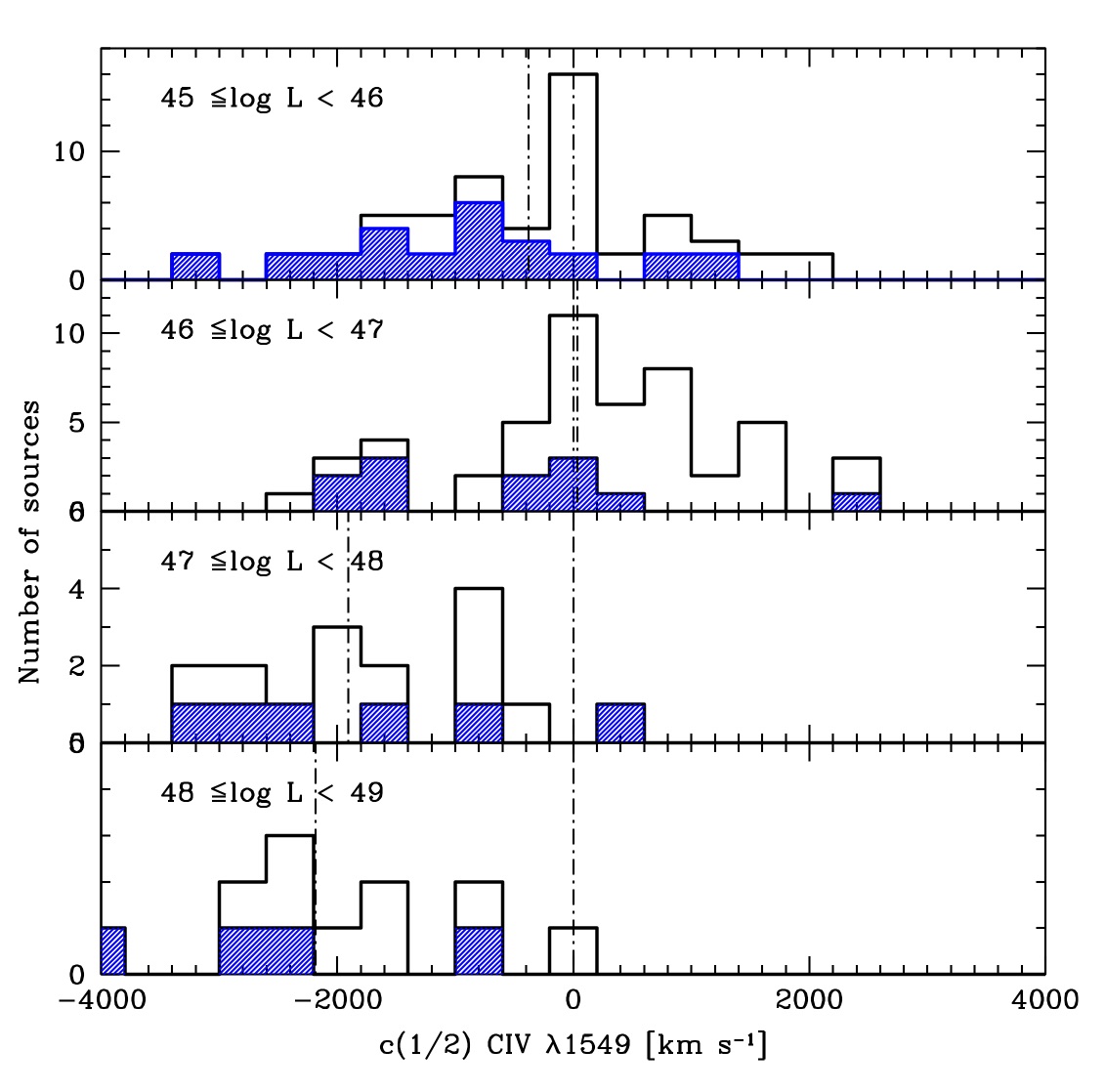} 
\includegraphics[scale=0.21]{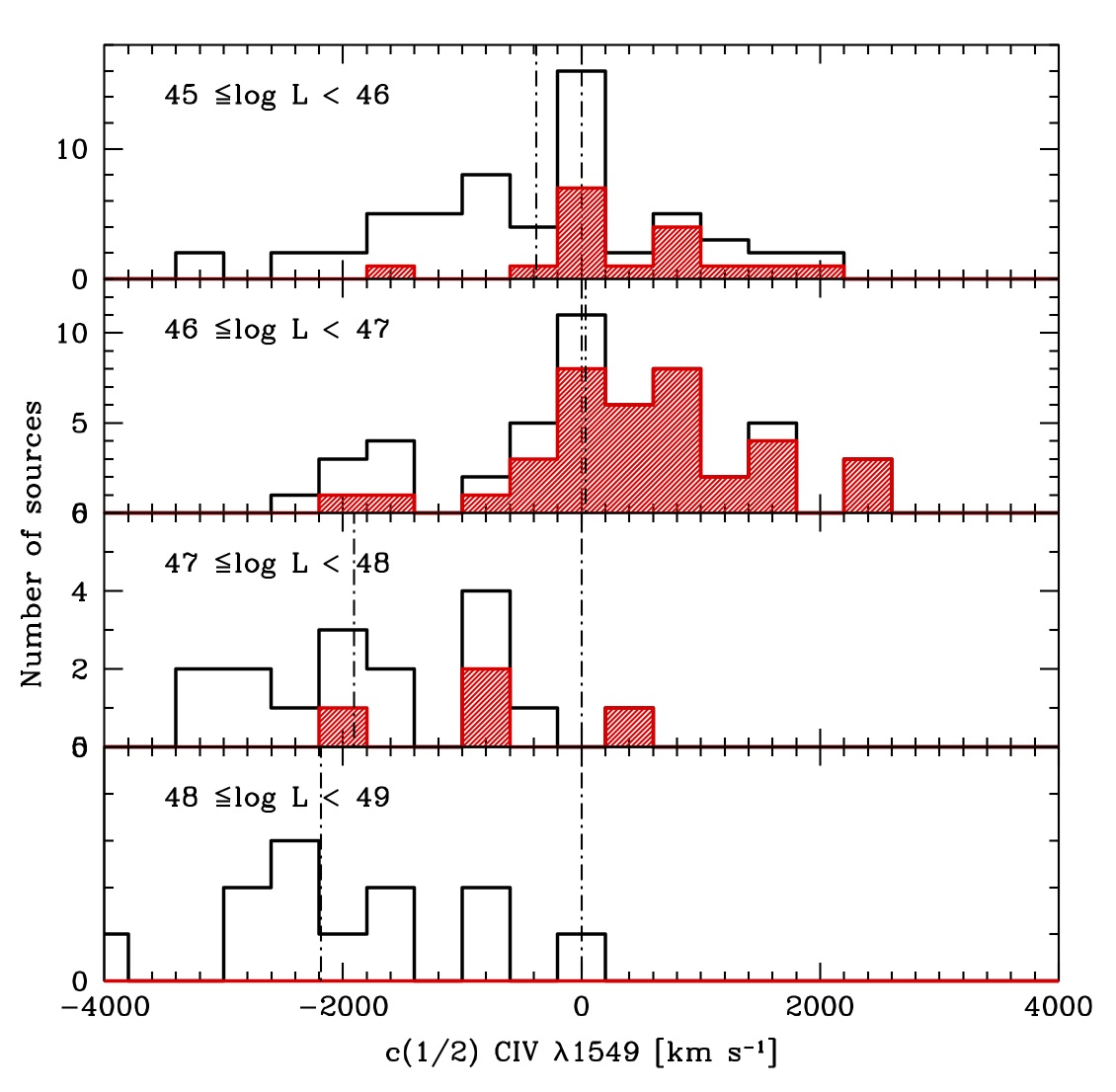}
\caption{Distribution of \civ\ centroids at half maximum, in \kms, in four quasar luminosity ranges, for the CS  and HE samples. Left:  the distribution of Population A sources is  shaded blue; right: RL distribution is shaded red. \label{fig:vdistr}        } 
\end{figure*}

\begin{figure}[htp!]
\includegraphics[scale=0.325]{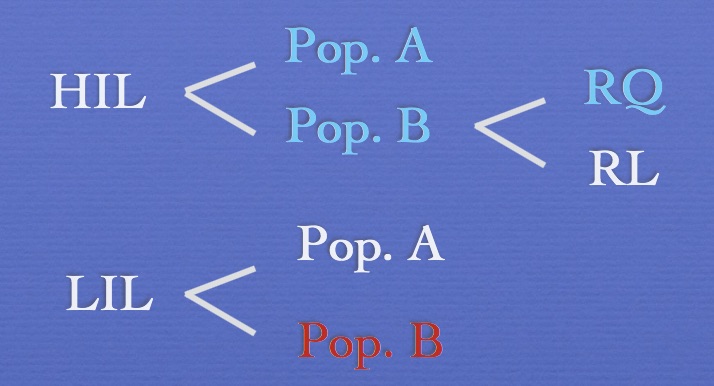}
\caption{Contextualization scheme for HILs and LILs. Separation of RL and RQ is needed  for HILs only in Pop. B. The color coding indicates a predominant blueshift (pale blue), no significant shift (white), and a predominant redshift (red). }
\label{fig:scheme}        
\end{figure}

The significance of these results can be appreciated by comparing the distribution of \civ\ line shifts (centroid at half maximum, $c(\frac{1}{2}$)) in four luminosity bins (Fig.  \ref{fig:vdistr}). The two panels for $\log L \ge$ 47 show the distribution for the HE sample, while the distributions at lower luminosity are drawn from the control sample. The left panel isolates Pop. A (shaded blue) while the right one RL sources (shaded red).  Pop. A sources are associated with the largest shifts, while RL sources dominate the redshifted part of the distribution at low $z$. It is not the case that RL sources do not show any blue shift; however the blueshift amplitudes rarely exceed 1000 \kms, and the luminosity dependence is shallower than for RQ \citep{richardsetal11}. We can consider that the effect of radio  loudness on \hb\ is not remarkable, if the comparison is restricted to Pop. B sources, as appropriate since there are very few Pop. A RQ. The redward asymmetry seen in Pop. B RL is seen in Pop. B RQ as well \citep{marzianietal03b,zamfiretal10}. It does not seem that the redward asymmetry is a phenomenon exclusive to RL sources \citep{punsly10,punslyzhang11}, as it is most prominent right in the HE sample Pop. B sources that are mostly RQ. Therefore a proper contextualization may follow the scheme of Fig. \ref{fig:scheme}: while for LILs it is sufficient to separate Pop. A and B to avoid mixup of sources in different dynamical conditions, for HIL analysis it is necessary to distinguish between Pop. B RQ and RL. The color coding in Fig. \ref{fig:scheme} reflects the average \civ\ centroid shifts at half maximum in the sample of \citet{sulenticetal07}: $c(\frac{1}{2}) \approx -900$, --250, +70  \kms\ for Pop. A RQ, Pop. B RQ, and Pop. B RL respectively. The weakness of the blueshifts in RL sources makes the redward asymmetry in \civ\ more easily detectable, so that the RL \civ\ profile either look more symmetric or even redshifted (Fig. \ref{fig:vdistr}), while the \civ\ profiles of RQ Pop. B sources are more affected by blue shifted emission, resulting in a smaller net blueshift on average. 

\begin{figure*}[htp!]
\includegraphics[scale=0.193]{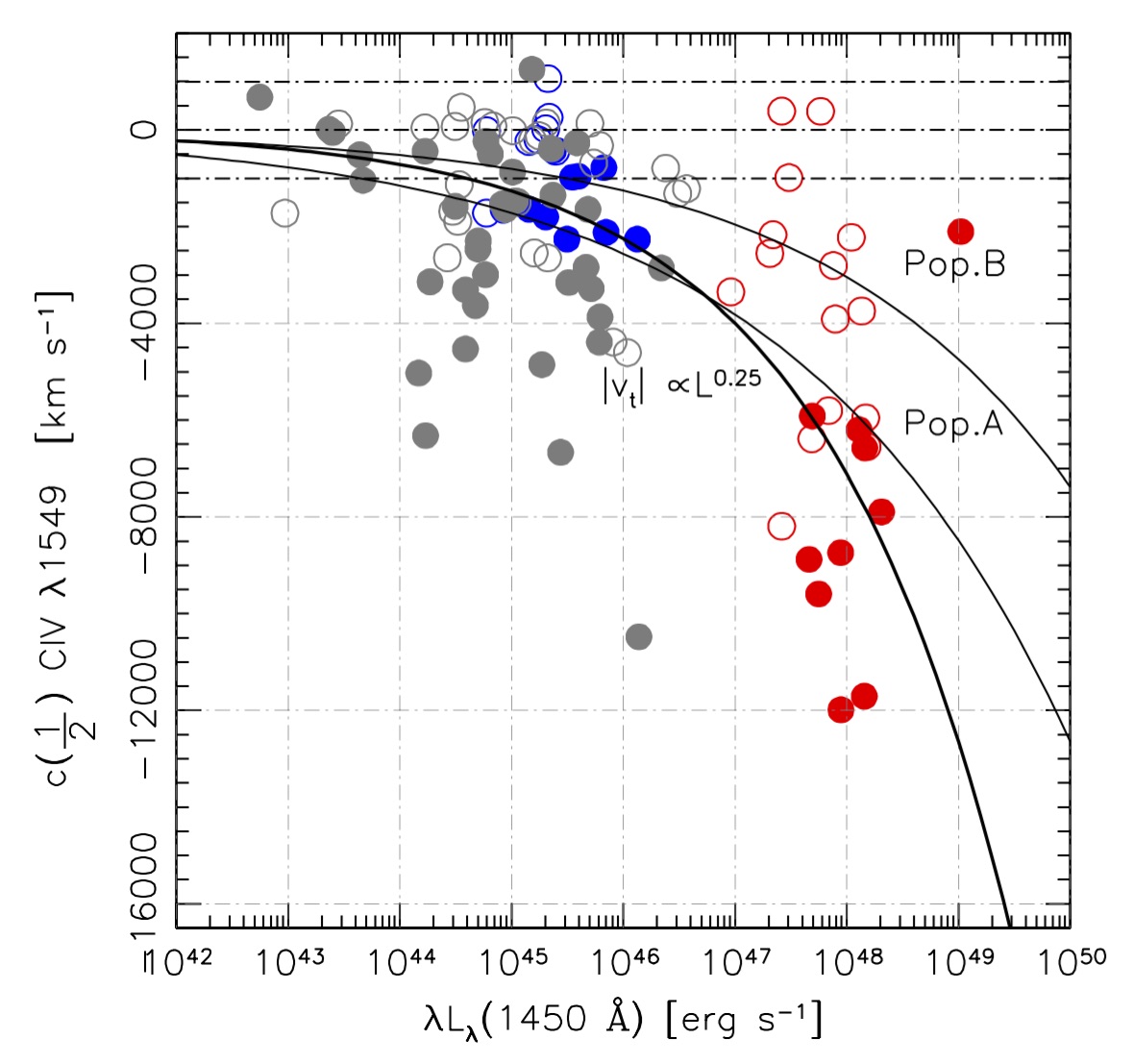} 
\includegraphics[scale=0.41]{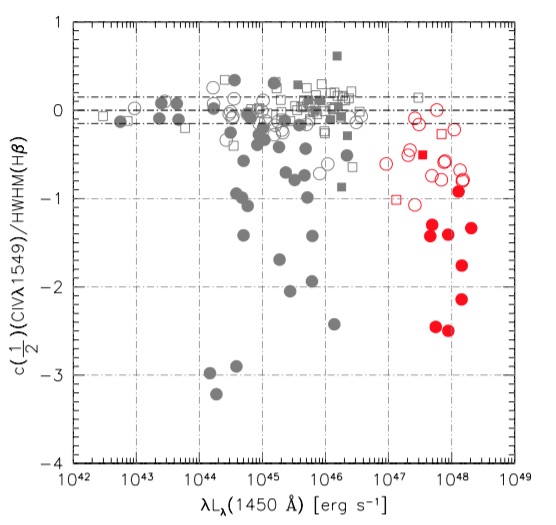}
\caption{Behaviour of the \civ\ emission line shifts as a function of continuum luminosity.  Left:  2 $c(\frac{1}{4})$ vs continuum luminosity at 1450 \AA. The thick black line shows the simple radiation-driven wind prediction, 2 $|c(\frac{1}{4})|\propto L^{\frac{1}{4}}$, the thin lines the best fit for Pop. A and B sources. Filled circles are Pop. A, open circles Pop. B. Data points refer to the CS (grey), to the HE sample (red) and to the faint high-$z$\ sample of \citet[][, blue]{sulenticetal14}. Right: dynamical relevance parameter $c(\frac{1}{2})|$/HWHM
(\hb) vs luminosity. Filled symbols are Pop. A; open Pop. B; circles RQ, squares RL. The dot-dashed lines limit the region where lines could be considered unshifted.   \label{fig:vlum}     }   
\end{figure*}

Fig.  \ref{fig:vdistr} indicates that blueshifts are more frequent at high luminosity, and that the shift amplitude tends to increase as a function of luminosity. This is also shown by Fig. \ref{fig:vlum}: the blueshift amplitude at the line base depends strongly on continuum luminosity, although it is probably misleading to interpret the diagram in terms of a correlation: at each luminosity there is a large range in shifts. In Fig. \ref{fig:vlum} we considered  twice the $c(\frac{1}{4})$\ as a proxy (as a matter of fact, an underestimate) of the terminal velocity $v_\mathrm{t}$\ associated with a wind. In the case of a radiation driven wind, $v_\mathrm{t}$ can be written as:

\begin{equation}
v_\mathrm{t} \sim v_\mathrm{K} \sqrt{{\cal M} \frac{L}{L_\mathrm{Edd}}}  \sim  \sqrt{\frac{GM}{L^{\frac{1}{2}}} }\sqrt{{\cal M} \frac{L}{M}}
\end{equation}
where we have assumed that the Keplerian velocity can be written as $v^{2}_{K} = GM/R$, and that the radius scales with the square root of the luminosity, as found from reverberation mapped sources \citep{bentzetal06,bentzetal09}. If the force multiplier ${\cal M}$ does not depend on $L$, then $v_\mathrm{t} \propto L^{\frac{1}{4}}$. In other words, in the case of a radiation driven wind, under the simplest assumptions, we expect  a pure luminosity dependence \citep[][c.f. \citealt{marzianietal13}]{laorbrandt02}. The line drawn in Fig. \ref{fig:vlum} for an arbitrary normalisation shows that a trend $|v_\mathrm{t}| \sim 2c(\frac{1}{4})  \propto L^{\frac{1}{4}}$\ is consistent with the systematic increase in Pop. A blueshifts (a different normalisation would account for Pop. B sources as well). 

 The dynamical relevance of the \civ\ shift i.e., the shift amplitude normalized by the line half-width at half-maximum of \hb\ \citep{marzianietal13a}, is $|\Delta v(\frac{1}{2})|$\-/HWHM(\hb) $ \gtsim$ 1 in the most extreme quasars. As is possible to appreciate from  Fig. \ref{fig:profiles}, the \civ\ line is almost blueshifted with respect to the  rest frame of the quasar. However, we do not see a strong dependence on line luminosity: the largest normalised shifts are observed in the control sample, at moderate $L$ ($\log L \sim 44$; right panel of Fig. \ref{fig:vlum}).



\subsection{Extreme outflows in extreme quasars: weak lined quasars}
\label{wlq}

\begin{figure}[htp]
\includegraphics[scale=0.19]{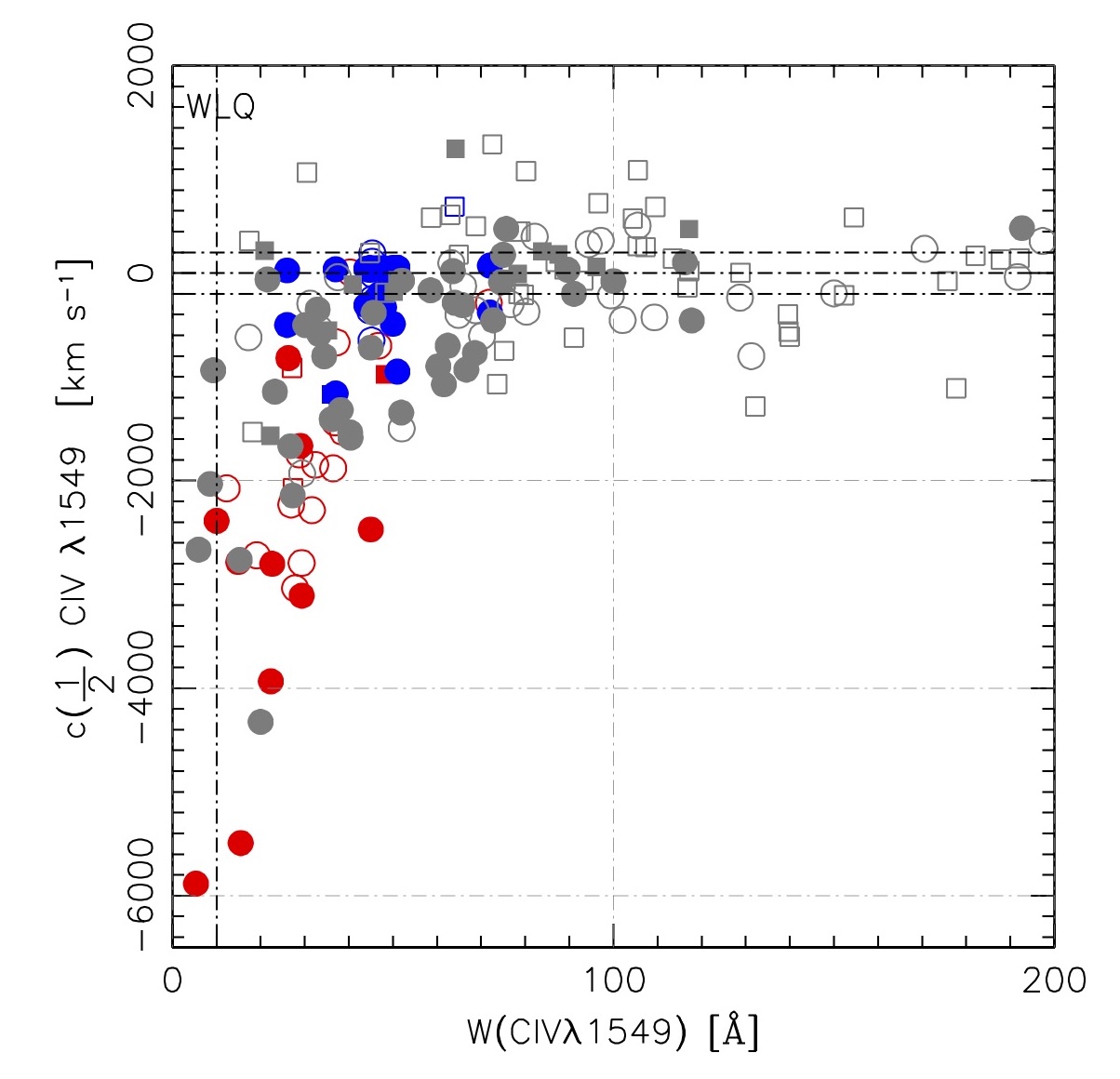} 
\caption{\civ\ shift in \kms\ vs rest frame W(\civ) in \AA. Filled symbols are Pop. A; open Pop. B; circles RQ, squares RL. The vertical dot-dashed line  limits the region of WLQs. Data points refer to the CS (grey), to the HE sample (red) and to the faint high-$z$\ sample of \citet[][, blue]{sulenticetal14}.}
\label{fig:vw}        
\end{figure}

\begin{figure}[htp!]
\includegraphics[scale=0.19]{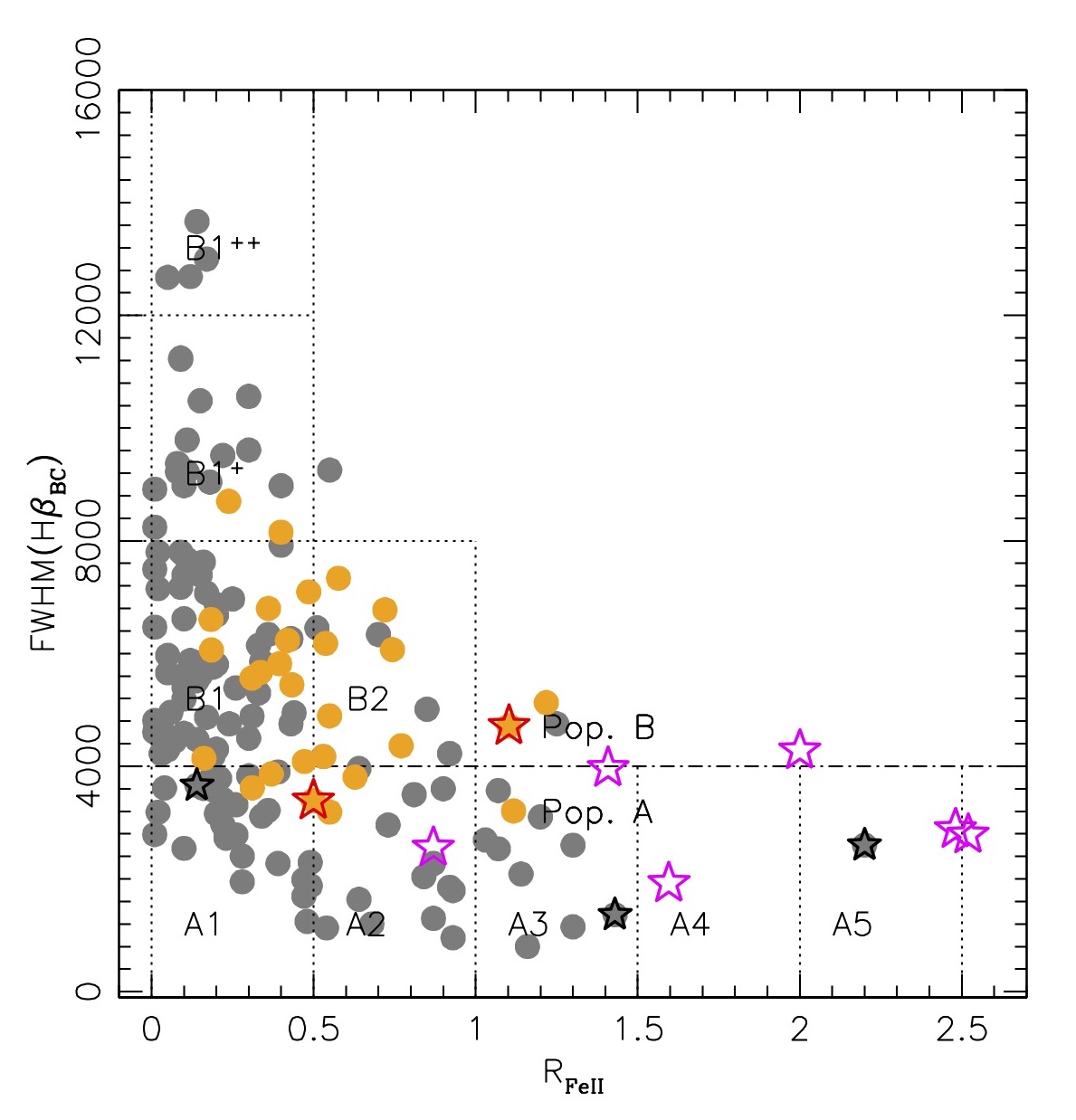} 
\caption{The optical 4DE1 plane, FWHM(\hb) vs. \rfe. Orange: HE data, grey: data from \citet{sulenticetal07}.  The stars identify the WLQs in the \citet{sulenticetal07} sample (black), in the HE sample (red),  and in the \citet{plotkinetal15} paper (magenta). }
\label{fig:e1}        
\end{figure}

Weak Lined Quasars (WLQs) have been a surprising discovery, since they are radio quiet quasars with usually low   equivalent width of \civ\ ($\le 10 $\AA) and \lya\ ($\le16$ \AA, \citealt{diamond-stanicetal09,shemmeretal10}). The insight gained by the contextualization of low redshift quasars \civ\ profiles allows to frame them relatively easily. Pop. A sources, in the 4DE1 context, show lower $W$(\civ) \citep{bachevetal04,sulenticetal07}, and among them we also found the most extreme blueshifts. 
We can consider at first the behaviour of the \civ\ shifts versus equivalent width (Fig \ref{fig:vw}). Of the five sources (3 in the CS and 2  HE) all low equivalent width sources show large \civ\ blueshifts ($\ltsim -1000$ \kms). The same results holds for the WLQs recently observed by \citet{plotkinetal15}, with simultaneous \hb\ and \civ\ coverage. Looking at the optical plane of the 4DE1 space (Fig. \ref{fig:e1}), we see that all 10 WLQs (CS, HE, and \citealt{plotkinetal15}) are Pop. A sources, and most of them  in the domain of extreme Population A with \rfe$\gtsim$ 1. Apart from the \civ\ line, other features in the UV spectra are  consistent with extreme A quasars revealed at both high and low luminosity, radiating at high Eddington ratio\ \citep{dultzinetal11,negreteetal12,marzianisulentic14}.  We conclude  that WLQs should be considered extreme Pop. A sources, probably among the quasars accreting at the highest rate. This view is supported by the optical and soft-X analysis of \citet{luoetal15} who found X-ray properties consistent with X-ray shielding by a thick structure possibly the ``slim'' disk expected at high accretion rate (\S \ref{4de1}).  

\section{Discussion:  what do the most powerful outflows  mean for galaxy evolution?}
\label{out}

The most powerful outflows are believed to produce significant feedback effects on the host galaxies  \citep{reevesetal09,rupkeveilleux11,fabian12}. The observational evidence of the active nucleus feedback is still debated, and rest upon the discovery of ultra-fast outflows (UFOs, \citealt{tombesietal10}), whose X-ray features are still passible of alternative interpretation. Large Balnicity index BAL QSOs are a minority of quasars \citep{vestergaard03,sulenticetal06a,trumpetal06}. On the converse, quasars outflows are observed in most radio quiet quasars (90\%\ of all quasars). An order of magnitude estimate of the kinetic power associated with the \civ\ emitting outflows is possible under several assumptions that make the estimate uncertain but not unrealistic \citep{fabian12,canodiazetal12,kingpounds15}. We start considering the fraction of the flux in the  \civ\ blue component that is above the expected projected escape velocity at $r \sim 1000$ \rg. The assumed $r$\ is relatively large in the BLR context and will yield a lower limit to the kinetic power.

\begin{figure*}[htp]
\includegraphics[scale=0.45]{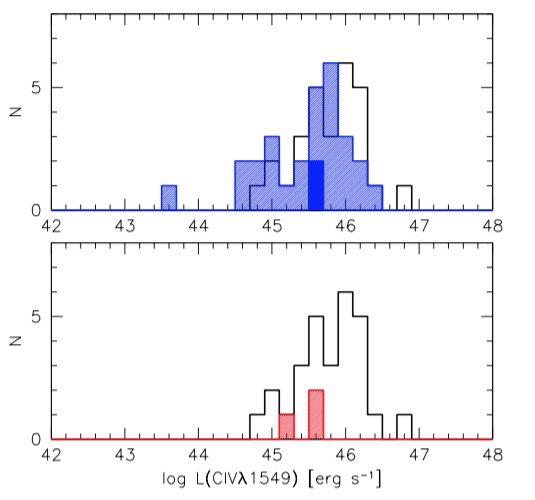} 
\includegraphics[scale=0.45]{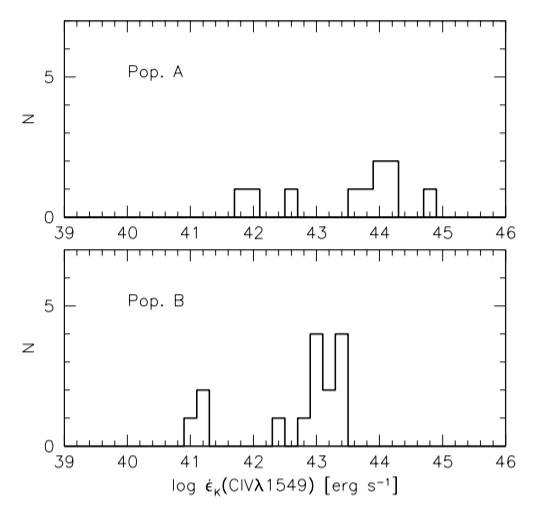}
\caption{Left: Distribution of \civ\ emission line luminosity, in \ergss. Top panel: shaded blue is the fraction above projected escape velocity. Dark blue identifies WLQs.  Bottom panel: same as top, with the identification of RL sources.  Right: distribution of kinetic power for Pop. A (top) and Pop. B (bottom), in units of \ergss.   }
\label{fig:ldistr}        
\end{figure*}

The \civ\  line luminosity associated with the outflow above escape velocity  is   given by
$$
 L({\rm CIV}) = \int _V j_\mathrm{CIV}~f_\mathrm{f}~dV
	 \label{eq:},$$
	 
	  (where $f_\mathrm{f}$\ is the filling  factor, and $j_\mathrm{CIV}$ is the line emissivity per unit volume) which can be written as: 
 
 $$j_\mathrm{CIV} = h \nu q_\mathrm{lu} n_\mathrm{e} n_\mathrm{l},$$
 
 for a collisionally excited line, where $n_\mathrm{e}$\ is the electron density and $n_{l}$\ the number density of C$^{3+}$\ ions at the transition lower level.  The collisional excitation rate $q_\mathrm{lu}$\   \citep{osterbrockferland06}Êis given by 
 
  $$q_\mathrm{lu} = \frac{\beta}{\sqrt{T}} \frac{\Upsilon_\mathrm{lu}}{g_{l}} \exp{\left(-\frac{\epsilon_\mathrm{lu}}{kT}\right)}$$  
  
where $g_\mathrm{l}$ is the statistical weight of the lower level, $\Upsilon_\mathrm{lu}$\ the effective collision strength, $T_\mathrm{e}$\ the electron temperature, $\epsilon_\mathrm{lu}$\ the energy level difference,  and $\beta$\ a constant.  The \civ\ line luminosity can be connected to the mass of outflowing ionized gas $M^\mathrm{ion}_\mathrm{out}$, by considering that the $M^\mathrm{ion}_\mathrm{out}$ can be written as, under the assumption of  constant density: 

$$
M^\mathrm{ion}_\mathrm{out} = 9.5~10^2~L_{45}({\rm CIV}) \left(\frac{Z}{5Z_{\odot}}\right)^{-1} n_{9}^{-1}~M_{\odot}
	 \label{eq_a5}
$$

where the metallicity $Z \approx 5 Z_{\odot}$\ is appropriate for luminous high $z$\ quasars \citep{hamannferland93,nagaoetal10}, and $n_{9}$ is the density in units of $10^{9}$ \cm3. The { mass outflow rate} at a distance $r$\ (1 pc) can be written, if the flow is confined to a solid angle of $\Omega$, as:    

$$ \dot{M}^\mathrm{ion}_\mathrm{out} =  \rho\ \Omega r^{2} v  = \frac{{M}^\mathrm{ion}_\mathrm{out}}{V} 	\Omega r^{2} v    \approx 15  L_{45} v_{5000} r^{-1}_{\rm 1} \rm M_{\odot}~yr^{-1} $$

The { outflow kinetic power}, with outflow $v$\ in units of 5000 km s$^{-1}$, is: 

$$ \dot{\epsilon} = \frac{1}{2} \dot{M}^\mathrm{ion}_\mathrm{out} v^{2} \approx 1.2 \cdot 10^{44} L_{45} v^{3}_{5000}   r^{-1}_{\rm 1} \rm  erg~s^{-1}. $$

The {  total energy  expelled over  a duty cycle} of 10$^{8}$ yr is 

$$ \int \dot{\epsilon} dt  \sim 4 \cdot 10^{59 }  L_{45} v^{3}_{5000}   r^{-1}_{\rm 1}  \tau_{8}  \rm~ erg.$$

This value can be compared to the { binding energy of the gas in a massive bulge/spheroid of mass $M_\mathrm{sph}$}:

$$ U = \frac{3 G M^{2}_\mathrm{sph}f_\mathrm{g}}{5 R_\mathrm{e}} \sim 2 \cdot 10^{59} M^{2}_\mathrm{sph,11} f_\mathrm{g,0.1} R^{-1}_\mathrm{e,2.5 kpc}\rm  erg,$$

 where $R_\mathrm{e}$\ is the effective radius in units of 2.5 kpc, and $f_\mathrm{g,0.1}$\ is the gas mass ratio in units of 0.1.
Fig. \ref{fig:ldistr} shows the distribution of the blue shifted component luminosity above the escape velocity at 1000 \rg\ (left panel). The distribution of the corresponding kinetic powers is shown in the right panel of Fig. \ref{fig:ldistr}, for Pop. A and B HE sources separately.  The luminosity values are enormous, comparable to the bolometric luminosity of bright quasars in the local Universe, and the kinetic energy deposed over a quasar life cycle is comparable to the binding energy of a massive spheroid.  We have assumed an escape radius of 1000 \rg, and density $n_{9} = 1$. Both parameters could likely have lower values, making our estimates lower limits to the $M^\mathrm{ion}_\mathrm{out}$\ and hence to the kinetic power.  Consistent results come from an analogous analysis of the \oiiiopt\ blue shifted emission. It is possible that we are, at least in part, considering the same outflows. However, if the \oiiiopt\ outflows can also be due to extranuclear star formation, this is very unlikely for the \civ\ broad line, that can be safely associated with processes occurring in the active nucleus. We also note that the WLQs do contribute significant kinetic power the low line equivalent widths notwithstanding, also because their line  blueshifts  are among the largest ever observed. 
 
\section{Conclusion} 
 
4DE1 provides an interpretation framework at low as well as at extreme luminosity. The \civ\ analysis reveals blueshifted emission associated with quasars outflows in RQ quasars, with high amplitude shifts being apparently more frequent at high $L$. Contextualization of HIL requires not only a distinction between Pop. A and B, but also that RL and RQ sources are kept separate, with RL sources showing blueshifts of systematically lower amplitude, no shifts or even redward asymmetric \civ\ profiles. 
``Unexpected'' luminosity effects are not seen (\S \ref{lumrl}): larger \civ\ shifts at high $L$\ are expected for a simple radiation-driven wind. Otherwise, the outflow phenomenology is self-similar over a wide range of $L$: Fig. \ref{fig:vlum} shows that large blueshifts ($\gtrsim 1000$ \kms) are possible over at least a 4.5 dex in luminosity, and that the \civ\ shift ``dynamical relevance'' is distributed over the same range at moderate and high $L$. The most extreme cases of blueshifts  include the WLQs that, in the 4DE1 context, are revealed to be predominantly  extreme accretions (i.e., WLQs belong to the  xA class of \citealt{marzianisulentic14}).  The \civ\ blueshift-dominated profile in the most luminous sources supports the idea that a nuclear outflow may be at the origin of galactic-scale feedback effects.
 
\begin{acknowledgements}
 We thank the referee, Martin Gaskell, for suggestions that improved the presentation of the paper, and a second anonymous reviewer for suggesting an important reference.  AdO and JS acknowledge the support by the Junta de Andaluc\' \i a through project TIC114,and the Spanish Ministry of Economy and Competitiveness (MINECO) through project AYA2013-42227-P.  PM acknowledges the hospitality of the Instituto de Astrof\'\i sica de Andaluc\' \i a (IAA-CSIC) where part of this work was done.  The TNG is operated on the island of La Palma by the Fundaci\'on Galileo Galilei of the INAF (Istituto Nazionale di Astrofisica) at the Spanish Observatorio del Roque de los Muchachos of the Instituto de Astrof\'\i sica de Canarias. 
\end{acknowledgements} 
%
%

\bibliographystyle{spr-mp-nameyear-cnd}

\bibliographystyle{spr-mp-nameyear-cnd}


\end{document}